\documentclass[a4paper,fleqn,usenatbib]{mnras}

\usepackage{newtxtext,newtxmath}

\usepackage[T1]{fontenc}
\usepackage{ae,aecompl}

\usepackage{graphicx}	
\usepackage{amsmath}	
\usepackage{amssymb}	
\usepackage{enumerate}

\title[Abell 1132]{LOFAR discovery of an ultra-steep radio halo and giant head-tail radio galaxy in Abell 1132}

\author[A. Wilber et al.]{
A. Wilber,$^{1}$\thanks{E-mail: amanda.wilber@hs.uni-hamburg.de (A. Wilber)}
M. Br{\"u}ggen,$^{1}$
A. Bonafede,$^{1,2}$
F. Savini,$^{1}$
T. Shimwell,$^{3}$
\newauthor
R. J. van Weeren,$^{4}$
D. Rafferty,$^{1}$
A. P. Mechev,$^{3}$
H. Intema,$^{3}$
F. Andrade-Santos,$^{4}$
\newauthor
A. O. Clarke,$^{5}$
E. K. Mahony,$^{6,7}$
R. Morganti,$^{8,9}$
I. Prandoni,$^{2}$
G. Brunetti,$^{2}$
\newauthor
H. R{\"o}ttgering,$^{3}$
S. Mandal,$^{3}$
F. de Gasperin,$^{3}$
M. Hoeft$^{10}$
\\
$^{1}$Hamburger Sternwarte, University of Hamburg, Gojenbergsweg 112, 21029 Hamburg, Germany\\
$^{2}$INAF/Istituto di Radioastronomia, Via P Gobetti 101, 40129 Bologna, Italy\\
$^{3}$Leiden University, Rapenburg 70, 2311 EZ Leiden, Netherlands\\
$^{4}$Harvard-Smithsonian Center for Astrophysics, 60 Garden Street, Cambridge, MA 02138, USA\\
$^{5}$University of Manchester, Jodrell Bank Centre for Astrophysics, Manchester, M139PL, UK\\
$^{6}$Sydney Institute for Astronomy, School of Physics A28, The University of Sydney, NSW 2006, Australia\\
$^{7}$ARC Centre of Excellence for All-Sky Astrophysics (CAASTRO)\\
$^{8}$ASTRON, the Netherlands Institute for Radio Astronomy, Postbus 2, 7990 AA, Dwingeloo, The Netherlands\\
$^{9}$Kapteyn Astronomical Institute, University of Groningen, P.O. Box 800, 9700 AV Groningen, The Netherlands\\
$^{10}$Thüringer Landessternwarte, Sternwarte 5, 07778 Tautenburg, Germany\\
}
\date{Accepted XXX. Received YYY; in original form ZZZ}

\pubyear{2017}

\begin{document}
\label{firstpage}
\pagerange{\pageref{firstpage}--\pageref{lastpage}}
\maketitle

\begin{abstract}

LOFAR observations at 144 MHz have revealed large-scale radio sources in the unrelaxed galaxy cluster Abell 1132. The cluster hosts diffuse radio emission on scales of $\sim$~650 kpc near the cluster center and a head-tail (HT) radio galaxy, extending up to 1 Mpc, south of the cluster center. The central diffuse radio emission is not seen in NVSS, FIRST, WENSS, nor in C \& D array VLA observations at 1.4 GHz, but is detected in our follow-up GMRT observations at 325 MHz. Using LOFAR and GMRT data, we determine the spectral index of the central diffuse emission to be $\alpha=-1.75\pm0.19$ ($S\propto\nu^{\alpha}$). We classify this emission as an ultra-steep spectrum radio halo and discuss the possible implications for the physical origin of radio halos. The HT radio galaxy shows narrow, collimated emission extending up to 1 Mpc and another 300 kpc of more diffuse, disturbed emission, giving a full projected linear size of 1.3 Mpc -- classifying it as a giant radio galaxy (GRG) and making it the longest HT found to date. The head of the GRG coincides with an elliptical galaxy (SDSS J105851.01$+$564308.5) belonging to Abell 1132. In our LOFAR image, there appears to be a connection between the radio halo and the GRG. The turbulence that may have produced the halo may have also affected the tail of the GRG. In turn, the GRG may have provided seed electrons for the radio halo.
\end{abstract}




\section{Introduction}

Cluster-scale diffuse radio emission, in the form of radio halos and radio relics, indicates the presence of large-scale magnetic fields and relativistic electrons within the intracluster medium (ICM). During a cluster merger, turbulence and shocks are produced in the ICM \citep[e.g.][]{2009MNRAS.395.1333V} and can lead to the re-acceleration of mildly-relativistic ICM electrons to ultra-relativistic speeds. The ultra-relativistic electrons (Lorentz factor $\gamma \gg 1000$) then interact with the ICM $B$-field (on the order of a few $\mu$G) to produce synchrotron emission in the radio regime, and can lead to the formation of large-scale radio sources called halos and relics (e.g. \citealp{2004rcfg.proc..335K}; see \citealp{feretti2012} for review). The origin of radio halos and relics involves complex mechanisms, and further investigation is needed to understand how these mechanisms affect the physics of the ICM (see \citealp{2014IJMPD..2330007B} for review). \\

Radio halos are classified as diffuse radio emitters that fill the central regions of galaxy clusters, and are found to coincide with the thermal gas seen in X-ray observations. These radio structures are vast in size, usually extending up to 1 Mpc, and are typically characterised by a steep spectrum ($\alpha \lesssim -1)$\footnote{We define the spectral index, $\alpha$, where $S\propto\nu^{\alpha}$.} and low surface brightness ($\sim$ 1 $\mu$Jy arcsec$^{-2}$ at 1.4 GHz; \citealp[e.g][]{feretti2012}). Two main models have been proposed for the origins of radio halos: the hadronic model and the turbulent re-acceleration model. The hadronic model states that collisions between cosmic-ray protons and thermal protons in the ICM would continuously produce the secondary electrons needed to generate radio halo emission at the cluster center \citep{1980ApJ...239L..93D, 1999APh....12..169B, 2011A&A...527A..99E}. Hadronic collisions should also produce secondary gamma-ray photons, but the non-detection of galaxy clusters in the gamma-ray regime has constrained the contribution from secondary electrons to be subdominant \citep[e.g.][]{2010ApJ...717L..71A, 2016ApJ...819..149A, 2011ApJ...728...53J, 2012MNRAS.426..956B}. The turbulent re-acceleration model states that mildy-relativistic ICM electrons are re-accelerated to ultra-relativistic energies \textit{in situ} during a cluster-sub-cluster merger \citep{brunetti2001, petrosian2001}. \\

Cluster mergers are thought to produce turbulence that can accelerate cosmic rays and may amplify the magnetic fields in the ICM via the small-scale dynamo \citep{2017MNRAS.464..210V, 2010ApJ...719L..74K, 2008Sci...320..909R, 2016IAUFM..29B.700M}. Indeed, studies that combine radio and X-ray data of galaxy clusters have suggested a causal link between merging activity of clusters and the occurrence of radio halos \citep[e.g.][]{2013ApJ...777..141C}. Currently, turbulent re-acceleration is the favoured scenario for the origin of radio halos, although several questions remain in identifying the seed source of mildy-relativistic electrons within the ICM and in understanding the physics of cluster shocks and shock-induced turbulence. There are also a few outliers, where giant radio halos are found in cool-core, non-merging clusters \citep[e.g.][]{2014MNRAS.444L..44B, 2017MNRAS.466..996S}, that challenge our present interpretation of ICM acceleration mechanisms. \\

A unique prediction of turbulent re-acceleration models is the existence of a large number of radio halos with very steep spectra \citep{2006MNRAS.369.1577C, 2008Natur.455..944B}. Steep-spectrum halos are produced when the turbulent re-acceleration rate is not efficient enough to accelerate electrons emitting at GHz frequencies, or during late evolutionary stages when turbulence is dissipated in the ICM \citep[e.g.][]{2012A&A...548A.100C, 2013MNRAS.429.3564D}. Searching for fading ultra-steep spectrum halos may assist in clarifying the physical origins of radio halos: the identification of breaks in the spectrum of halos can be used to infer the efficiency of the mechanism that produces the emitting cosmic-ray electrons \citep{2003A&A...397...53T, 2010MNRAS.401...47D, 2010MNRAS.407.1565D}. A few cases of ultra-steep spectrum halos have been found \citep[e.g.][]{2008Natur.455..944B, 2013A&A...551A.141M, 2012MNRAS.426...40B}, but further sensitive low-frequency observations may be needed to reveal the population of ultra-steep halos. \\

Active galactic nuclei (AGN) injection from individual cluster radio galaxies is one explanation for the large supply of mildly-relativistic seed electrons needed for the turbulent re-acceleration that produces cluster-scale radio emission (such a connection has been established for certain cluster radio relics, as for example in \citet{2017NatAs...1E...5V}). Giant radio galaxies (GRGs), which are generally defined as radio galaxies with a linear projected size of $\gtrsim 0.7$ Mpc \citep[e.g.][]{2005AJ....130..896S}, can have a significant influence on their surrounding medium by supplying a large quantity of cosmic rays. Different scenarios may explain the formation and large extent of GRGs: the AGN may have been active for a very long time, the jets may be powerful enough to push emission out to large distances without much deterrence from the surrounding medium, and/or the surrounding medium could be much less dense as compared to the medium around typical radio galaxies \citep{Kaiser1999}.\\

Tailed and bent-tailed radio galaxies are often found within the rich environments of galaxy clusters, and their jets give an indication of where cosmic-ray electrons are being injected into the surrounding medium. These tailed sources have been typically categorized by their morphologies as seen in projection: as wide-angle-tail (WAT) when two radio jets, or plumes, are distinguishable, collimated, and open at an angle of $\lesssim$ 60$^{\circ}$, or narrow-angle-tail (NAT) when the radio jets open in a very small angle such that they appear aligned on one side of the host galaxy or conjoined as a single tail (also referred to as head-tail (HT) radio galaxies) \citep[e.g.][]{2014AJ....148...75D}. It is generally thought that bent-tailed galaxies form when the jets experience ram pressure as the host galaxy moves through the ICM. A host galaxy traveling at high velocity may experience a ram pressure shock that aligns both radio jets behind the host's trajectory, leaving the radio source with a perceived NAT/HT morphology \citep{1980ARA&A..18..165M}. Of the known bent-tailed radio galaxies, only a small percentage are also GRGs. The longest HT discovered so far is in Abell 1314 with a projected linear size of 700 kpc \citep{Srivastava2016}. \\

\subsection{Abell 1132}

Abell 1132 is a massive cluster ($5.87^{+0.22}_{-0.23} \times 10^{14}$M$_{\sun}$ from the \citealp{2014A&A...571A..29P}) that shows signs of merging \citep{2015A&A...580A..97C}, but has not shown diffuse radio emission in past VLA observations at 1.4 GHz \citep{Giovannini2000}. Abell 1132 is centred at R.A., decl. 10h58m25.8s, +56$^{\circ}$47$\arcmin$30$\arcsec$ (equatorial, J2000.0) and located at a redshift of $z = 0.1369$ \citep{1999ApJS..125...35S}. It contains several Fanaroff-Riley (FR) type-I radio galaxies \citep{O'Dea1985}, was covered by the NRAO VLA Sky Survey (NVSS; \citealp{condon1998}), and has been observed by the Chandra X-ray Observatory.  \\

\citet{Rudnick2009} reported an extended head-tail source, 370 kpc long, about 6$'$ south of the cluster center using reprocessed data from the Westerbork Northern Sky Survey (WENSS; \citealp{Rengelink1997}). They noted that the head of this source is visible in NVSS and coincides with an elliptical galaxy\footnote{\label{opticalgalaxyHT} mR = 16.78 galaxy in SDSS (R.A., decl. = 10h58m50.96s, +56$^{\circ}$43\arcmin08\arcsec), with a redshift of z = 0.138954 $\pm$ 0.000162.} belonging to the cluster. \\

In this paper, we report on Low-Frequency Array (LOFAR) observations and follow-up Giant Meterwave Radio Telescope (GMRT) observations of the galaxy cluster Abell 1132 and newly discovered extended radio emission. In the following section, details of our observations, data calibration, and imaging techniques are described. In Sec.~\ref{results} we show our LOFAR and GMRT images of the radio emission seen in Abell 1132, and in Sec.~\ref{discuss} we discuss how these images provide clues as to the morphology and possible origins of the detected emission. The scale at Abell 1132's redshift is 2.439 kpc$\arcsec^{-1}$ with the cosmological parameters H$_{0}=69.6$, $\Omega_{m}=0.286$, and $\Omega_{\Lambda}=0.714$, adopted hereafter. 

\section{Methods}

\subsection{LoTSS}
LOFAR is a low-frequency radio interferometer based in the Netherlands with additional international stations throughout Europe \citep{vanHaarlem2013}. The array includes low-band and high-band antennae (LBA and HBA) that receive signals over a frequency range of 10-90 MHz and 120-240 MHz, respectively. LOFAR's large field-of-view, compact core, and its high sensitivity at low frequencies makes it the ideal instrument to study steep spectrum diffuse radio emission in galaxy clusters as well as giant radio galaxies with large angular diameter and low surface brightness. \\

The LOFAR Two-meter Sky Survey (LoTSS) \citep{2017A&A...598A.104S} Tier-1 has currently observed 14\% of the northern sky at 120 -168 MHz (as of July 2017) with one of its main science goals to study galaxy clusters and search for cluster-scale radio emission. With a depth of $\sim 100$~$\mu$Jy beam$^{-1}$ at a resolution of $\sim 5 ''$, this survey is particularly sensitive to steep spectrum radio emission, and thanks to the compact core of LOFAR it can detect radio objects with low surface brightness \citep[see][for details]{shimwell2016, 2017A&A...598A.104S}.  \\

The observation of Abell 1132 was part of a standard Tier-1 survey observation, covering an area of $\sim$~19 deg$^{2}$ centered on R.A., decl. 11h00m20s, +57$^{\circ}$11\arcmin48\arcsec, made with the LOFAR Dutch HBA array\footnote{The Dutch array consists of 24 core and 14 remote stations in the Netherlands.}. The north-western region of this field overlaps the Lockman hole (a region of space that has a low column density of hydrogen) which was also covered by LoTSS. Although Abell 1132 is visible in those observations (see \citealp{mahony2016} and \citealp{2017arXiv170801904B} for details), here we provide a more detailed analysis of the radio emission seen in Abell 1132 with our more recent and more sensitive Tier-1 observation centered on Abell 1132. Our survey field was observed for 8 h to ensure sufficient $uv$ coverage. The observation covers a total bandwidth of 48 MHz with a central frequency of 144 MHz.  See Table~\ref{obstable} for observation details. \\

\begin{table*}
 \centering
   \caption{Radio observations of A1132.} \label{obstable}
  \begin{tabular}{c c c}
  \hline
Telescope & LOFAR & GMRT  \\
Observation ID & 544905 & 9101 / 9104 \\
Pointing center (RA,DEC: J2000) & 11h00m20s, +57$^{\circ}$11\arcmin48\arcsec  & 10h58m25.8s   +56$^{\circ}$47\arcmin30\arcsec \\
Observation date & 2016 Sep 6 & 2016 Dec 30 / 31 \\
Total on-source time & 8 h & 8 h \\
Flux calibrator & 3C196 & 3C147 \& 3C286  \\
Total on-calibrator time & 10 min & 20 min\\
Central frequency & 144 MHz & 325 MHz / 610 MHz \\
Bandwidth  & 48 MHz & 16 MHz / 32 MHz \\
\hline
\hline
\end{tabular}
\end{table*}

\subsection{LOFAR data reduction}

The standard data reduction for LoTSS data can be summarized into the following steps:
\begin{itemize}

\item Pre Facet Calibration via Prefactor\footnote{\url{https://github.com/lofar-astron/prefactor}}: Computes direction-independent solutions from observation of a standard calibrator and transfers these solutions to the target data. Performs an initial phase calibration for the target data using a global sky model. Produces preliminary images of the full field of view.

\item Facet Calibration via FACTOR\footnote{\url{https://github.com/lofar-astron/factor}}\citep{vanWeeren16}: Performs direction-dependent calibration in multiple directions over the full bandwidth of target data. Corrects for ionospheric disturbances and beam errors. Produces high-resolution images.
\end{itemize}

In the following subsections, these steps will be elaborated upon for this particular observation.

\subsubsection{Pre Facet Calibration}

The calibrator chosen for this observation was 3C196, a bright quasar (74~Jy according to the \citet{2012MNRAS.423L..30S} absolute flux scale). The Prefactor pipeline was used to compute the amplitude gains, station clock offsets, station phase correlation offsets, and station differential total electron content (dTEC) from the standard calibrator data and apply all direction-independent solutions to the target data\footnote{dTEC solutions from the calibrator were not transferred to the target data since they correspond to the ionosphere and are therefore direction-dependent.}. After direction-independent solutions were transferred, an initial phase calibration was performed on the target field using a global sky model produced from the VLA Low-Frequency Sky Survey (VLSSr; \citealp{Lane2012}), WENSS \citep{Rengelink1997}, and NVSS \citep{condon1998}. Data from the station CS030HBA were discarded (flagged) because it was not operational for most of the observation. \\

The Initial-Subtract step, as part of the Pre Facet Calibration pipeline, imaged the direction-independent calibrated target data in both high and low resolution using {\tt WSClean} \citep{2014MNRAS.444..606O}. The pipeline first created an image at a resolution of $\sim36\arcsec\times27\arcsec$ and automatically detected and masked sources using the source detection software PYBDSF\footnote{\url{http://www.astron.nl/citt/pybdsf/}}. The clean components of the masked sources were subtracted from the $uv$-data. Then an image was created at a resolution of $\sim120\arcsec\times100\arcsec$ where more extended sources were revealed, masked, and also subtracted. The final result was a sky model with all high-resolution (compact sources) and low-resolution sources (extended sources) subtracted and a source-subtracted $uv$-dataset, which was then used for direction-dependent calibration. \\

\begin{figure*}
\centering
\includegraphics[width=\textwidth]{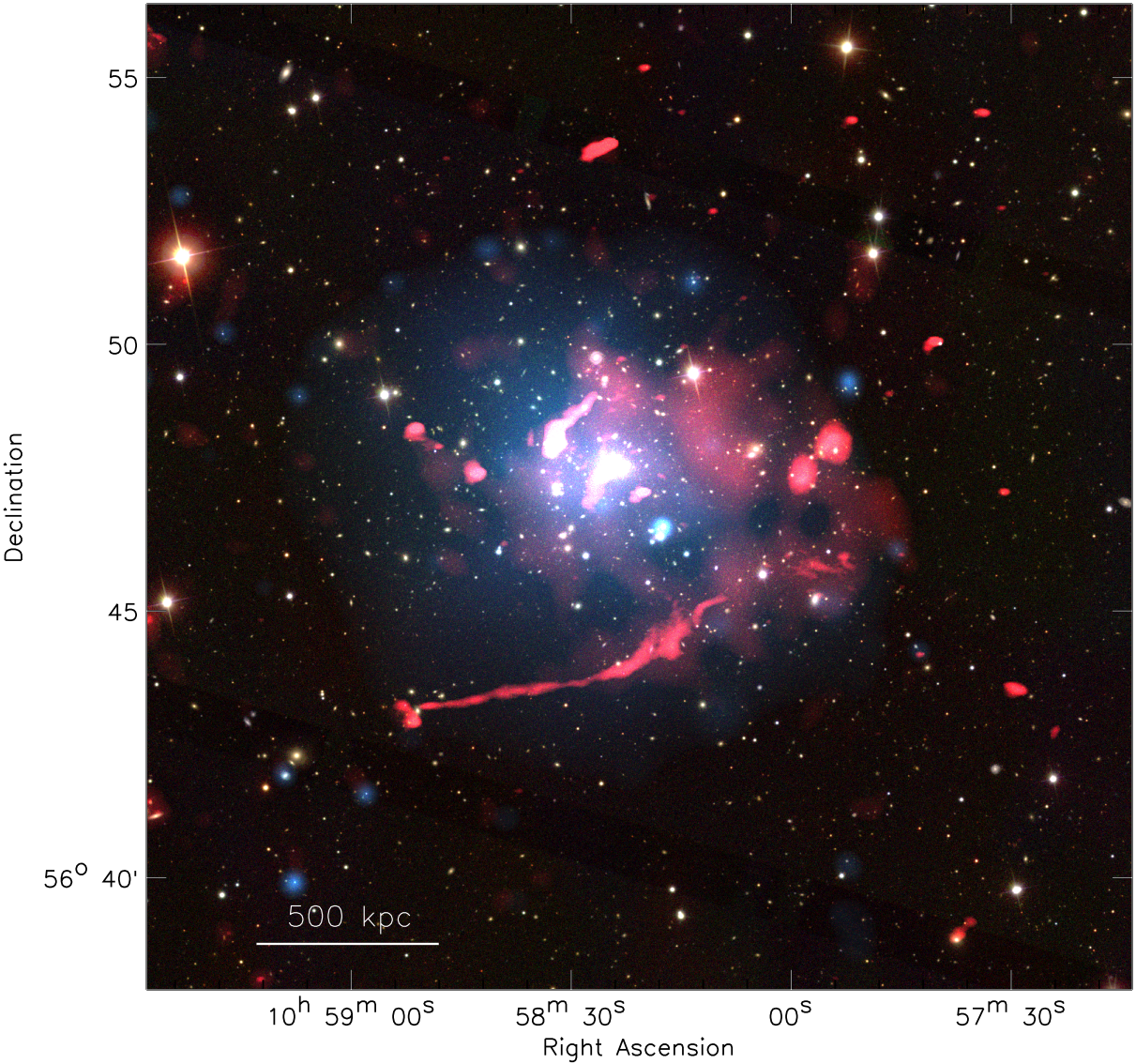}
\caption{Overview image of Abell 1132: LOFAR high-resolution and low-resolution emission are both shown in red. LOFAR high-resolution emission is imaged with a beam size of $\sim8\arcsec\times5\arcsec$ and RMS noise of $120~\mu$Jy beam$^{-1}$. LOFAR low-resolution diffuse emission (imaged after performing a subtraction of compact sources, as explained in Sec.~\ref{3.1}) is imaged with a beam size of $\sim30\arcsec\times26\arcsec$ and RMS noise of $350~\mu$Jy beam$^{-1}$. Chandra X-ray emission is in blue. Radio and X-ray emission are overlaid on optical SDSS \textit{g, r, i} images. Diffuse radio emission is present near the center of the cluster, as well as in the length of the southern HT GRG. The central diffuse radio emission, characteristic of a radio halo, is offset from the center of the X-ray emission by $\sim200$ kpc. There also appears to be a connection between the radio halo and the diffuse radio emission of the GRG tail. \label{Abell1132-pr}}
\end{figure*}

\subsubsection{Facet Calibration} \label{Facet Calibration}
Facet Calibration is a direction-dependent calibration method for LOFAR, implemented by the FACTOR\footnote{\url{http://www.astron.nl/citt/facet-doc/}} package \citep{vanWeeren16}. The first step is to tesselate the full field into multiple facets. Each facet must have its own calibrator, designated by a square region within the facet. The calibrator region is chosen by default to be centered on a compact source at least 0.3 Jy in flux density. The calibration region can be modified by the user to include multiple sources (increasing the calibrator flux density) and/or to include extended emission. To accurately calibrate a typical LoTSS widefield, up to 50 facet directions may be needed. \\

A process of self-calibration is conducted within the calibration region per facet, where several phase only and phase + amplitude self-calibration cycles are performed until there is convergence. The self-calibration solutions of a given calibrator region are then applied to its full facet, and the facet is imaged using {\tt WSClean}. Each subsequent facet is imaged where all sources from prior facets are subtracted. Typically, facets are processed in order of calibrator brightness; this gradually decreases the effective noise in the $uv$-data. Facet images can be stitched together and primary-beam-corrected for a mosaic image of the full target field. For more details on Facet Calibration the reader is referred to \citet{vanWeeren16}, \citet{shimwell2016}, and \citet{2016MNRAS.460.2385W}. \\

FACTOR was run on the full 48 MHz bandwidth for our observation. A total of 39 directions were designated and 15 bright and nearby facets were processed before processing the target facet containing the cluster Abell 1132. Initially our target calibration region included Abell 1132's complex cluster center. The calibration in this region was inadequate since emission was being displaced and creating bright and negative artifacts. We decided instead to use the calibration solutions from the nearest facet, $\sim0.25^{\circ}$ east of Abell 1132's center. The final image produced by FACTOR using {\tt WSClean} has a beam size of $\sim8\arcsec\times5\arcsec$ with root mean square (RMS) noise of $\sigma\approx120~\mu$Jy beam$^{-1}$.  \\

The FACTOR-calibrated data were also imaged outside of FACTOR using CASA (Common Astronomy Software Applications; \citealp{2007ASPC..376..127M}) tools. CASA {\tt CLEAN} was used with various adjusted parameters ($uv$-taper and Briggs' robust\footnote{\url{http://www.aoc.nrao.edu/dissertations/dbriggs/}} weighting schemes) so that diffuse emission would be properly masked and deconvolved. An increased outer $uv$-taper was used to bring out diffuse emission at lower resolutions. 

\subsection{Chandra X-ray data reduction}

Abell 1132 was observed with the Chandra ACIS-I (ObsID: 13376) in Aug 2011 for 8 ks. We processed the Chandra data following \citealp{Vikhlinin2005}\footnote{We used CIAO v4.6 and CALDB v4.7.2.}. This processing includes filtering of periods with elevated background by examining the light curves in the 6--12~keV band, the application of gain maps to calibrate photon energies, and corrections for the position-dependent charge transfer inefficiency. For the final exposure corrected image we used a pixel binning of 4 (2\arcsec pixel$^{-1}$). The instrumental and sky background were subtracted. For more details the reader is referred to \citet{Vikhlinin2005}. \\

To obtain the global temperature and luminosity within $R_{500}$ we fitted the spectrum in {\tt XSPEC} \citep[v12.9,][]{Arnaud1996}, extracting counts in the 0.7--7.0 keV band. For $R_{500}$ we took a value 1.218 Mpc, derived from the mass of $5.9 \times 10^{14}$ M$_{\sun}$ \citep{2014A&A...571A..29P}. Compact sources were excluded from the fitting. The abundance was fixed to 0.3 Z$_{\sun}$ \citep[from the abundance table of][]{1989GeCoA..53..197A} and the redshift at $z = 0.1366$. The hydrogen column density, N$_{\rm{H}}$, was fixed to $6.29 \times 10^{19}$cm$^{-2}$, adopting the galactic value (atomic + molecular) from \cite{2013MNRAS.431..394W}. \\

From the  {\tt XSPEC} fitting we determine a global temperature of $7.85^{+0.46}_{-0.46}$ keV and a luminosity of $(4.4\pm0.1)\times 10^{44}$~erg~s$^{-1}$ in the 0.1--2.4 keV energy band and $(2.7\pm0.1)\times 10^{44}$~erg~s$^{-1}$ in the 0.5--2.0~keV energy band. The bolometric luminosity is $(10.6\pm0.3)\times 10^{44}$~erg~s$^{-1}$. \\

Abell 1132 is an unrelaxed, merging cluster according to its disturbed X-ray morphology revealed by Chandra, and it is one of the few merging systems belonging to samples of massive clusters without diffuse radio emission seen at higher frequencies \citep{2015A&A...580A..97C}. Given the $L_{1} - T_{1}$ relation (BCES (Y$\vert$X) fitting) in Table 2 of \citet{2009A&A...498..361P}, the global temperature we measure ($7.85^{+0.46}_{-0.46}$ keV) would correspond to a luminosity of $(2.2\pm0.1)\times 10^{45}$~erg~s$^{-1}$ in the 0.1--2.4 keV energy band. This is an order of magnitude higher than our measured luminosity, thus fitting into the evidence that suggests this cluster is a merger. \\

\subsection{GMRT data reduction}
Follow-up GMRT observations of Abell 1132 at 325 MHz and 610 MHz were performed on December 30, and 31, 2016 (see Table \ref{obstable} for observation details). The GMRT data were processed using the SPAM pipeline \citep[see][for details]{2017A&A...598A..78I}. Our images with the highest resolution and lowest noise were produced in AIPS (Astronomical Image Processing System; \citealp{Wells1985}). These data were also imaged with CASA {\tt CLEAN} interactively with various adjusted parameters so that diffuse emission would be properly masked and de-convolved.  

\begin{figure*}
\centering
\includegraphics[width=\textwidth]{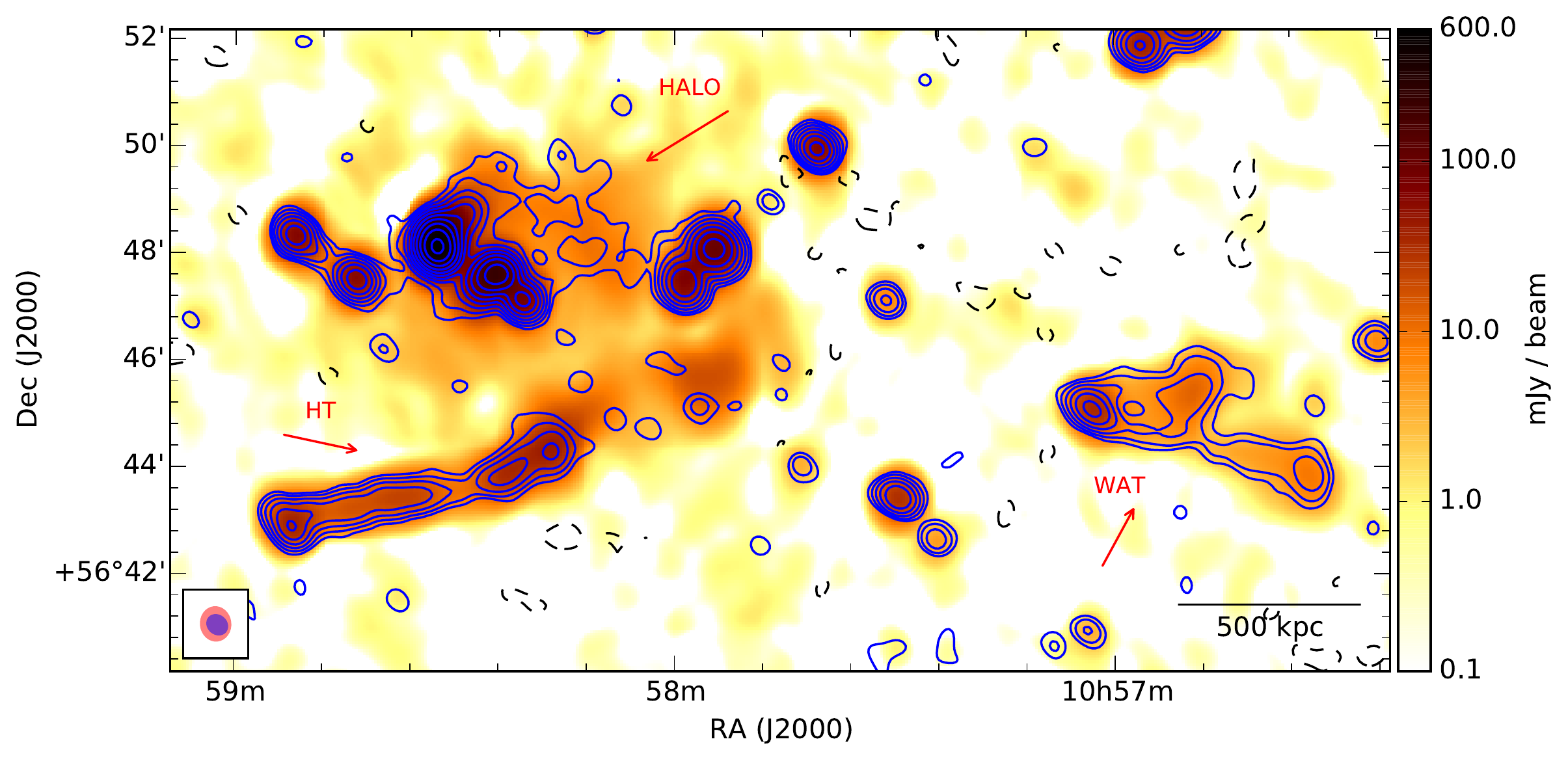}
\caption{Abell 1132: 144 MHz LOFAR low-resolution emission is shown in color (on a logarithmic scale) with 325 MHz GMRT low-resolution contours overlaid in blue. LOFAR and GMRT emission are imaged in CASA {\tt CLEAN} with $uv$-taper 25$''$ and Briggs' robust 0, with respective RMS noise of $400~\mu$Jy beam$^{-1}$ and $220~\mu$Jy beam$^{-1}$. Contours are $\sigma\times[-2, 2, 4, 8, 16, 32, 128, 256, 512, 1024]$. Beam size is designated by the red (LOFAR $\sim40\arcsec\times35\arcsec$) and blue (GMRT $\sim26\arcsec\times22\arcsec$) ellipses. Diffuse emission is present near the cluster center as well as in the westward portion of the giant HT. A WAT is visible $\sim 0.2^{\circ}$ (1.8 Mpc) west of the cluster center. \label{lo-contours}}
\end{figure*}

\section{Results} \label{results}

\begin{figure*}
  \centering
    \includegraphics[width=0.49\textwidth]{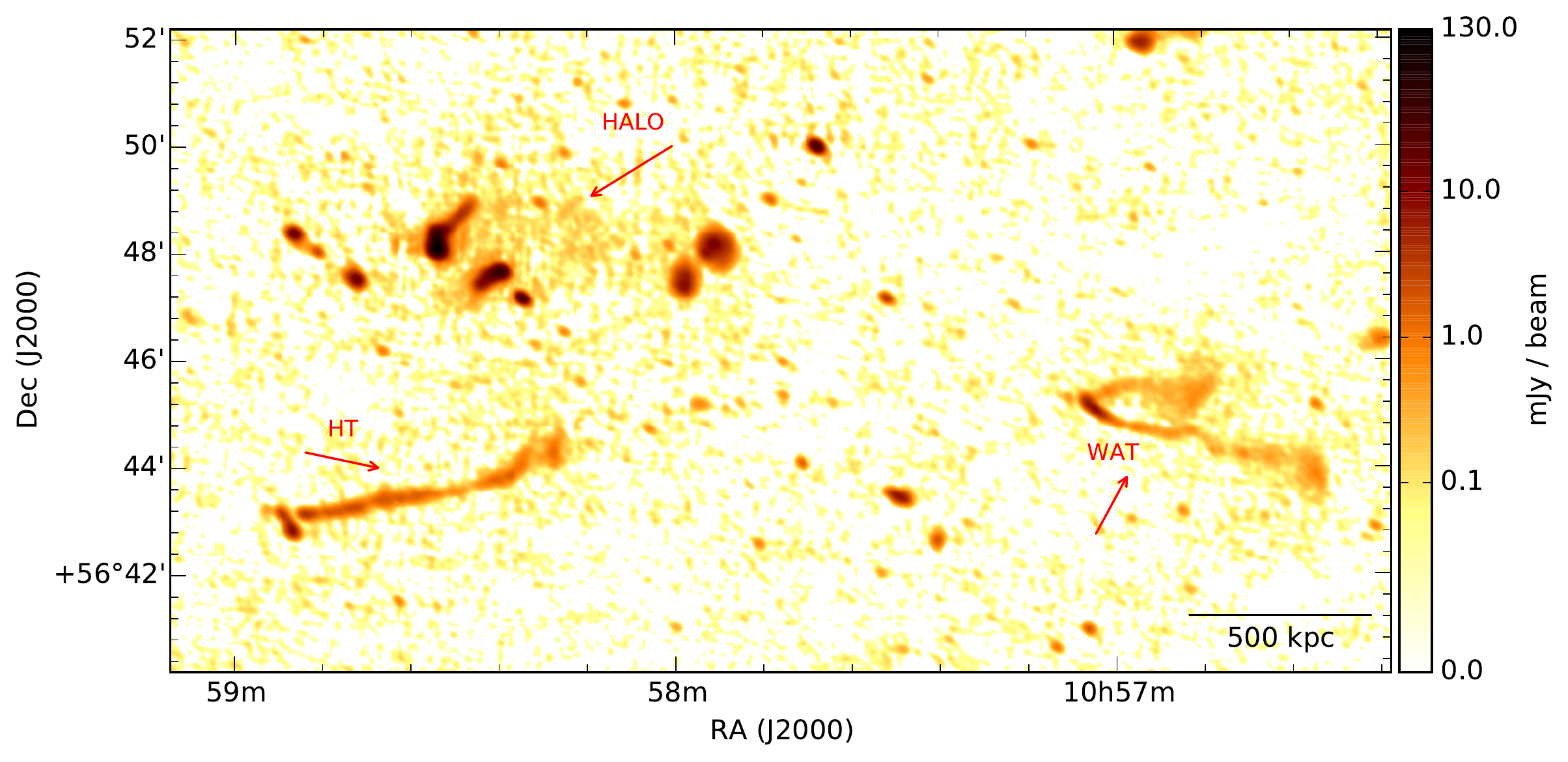}
    \includegraphics[width=0.49\textwidth]{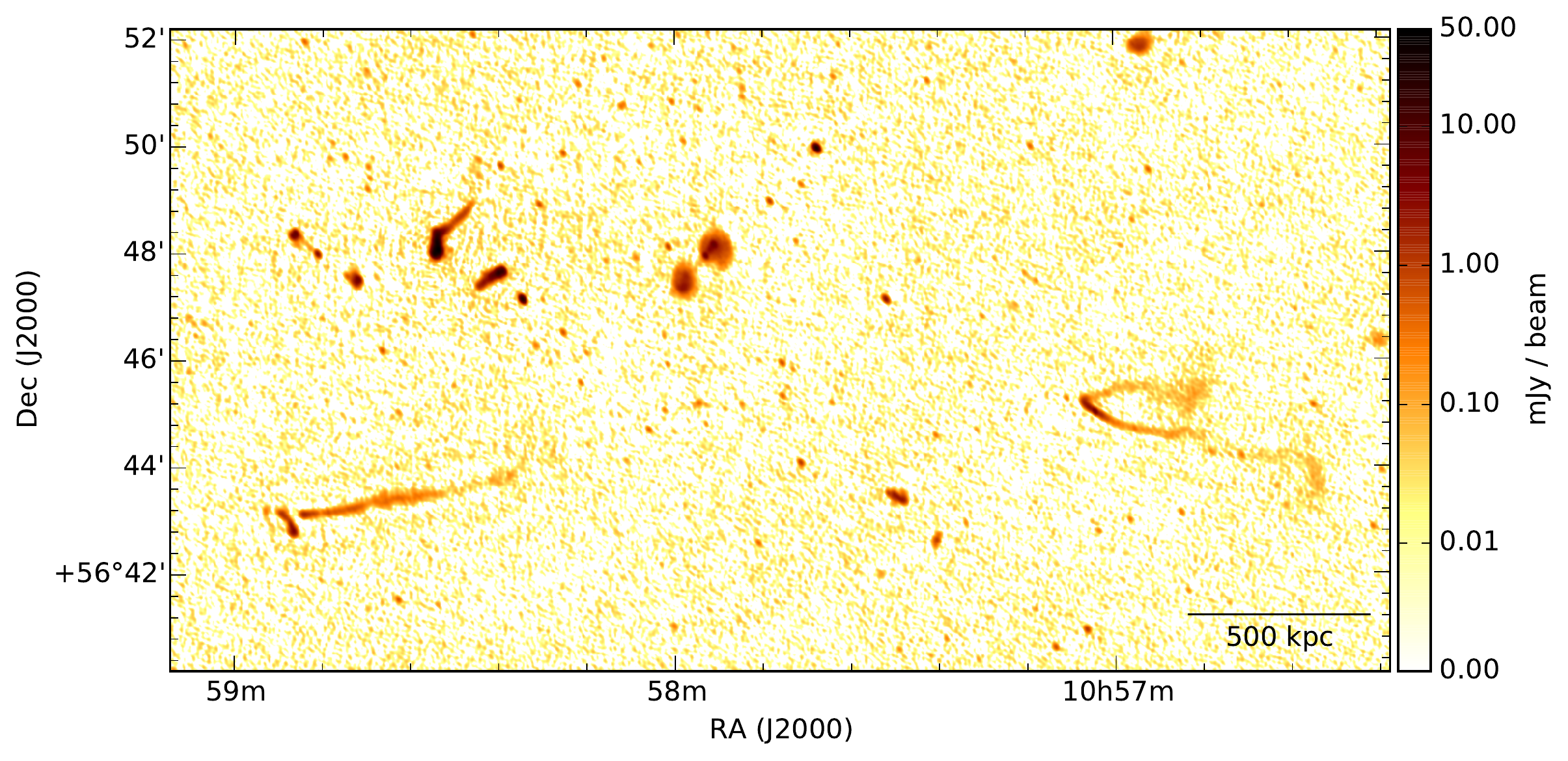}
    \caption{Left: GMRT image at 325 MHz with a resolution of $\sim10\arcsec\times7\arcsec$ and RMS noise of $\sigma \approx 45~\mu$Jy beam$^{-1}$. Right: GMRT image at 610 MHz with a resolution of $\sim6\arcsec\times4\arcsec$ and RMS noise of $\sigma \approx 20~\mu$Jy beam$^{-1}$. GMRT data was calibrated via the SPAM pipeline and imaged in AIPS with Briggs' robust -1. Both GMRT images show a significant portion of the HT radio galaxy and the full WAT radio galaxy. The radio halo is partially visible at 325 MHz. \label{325}}
  \end{figure*}

\begin{figure*}
  \includegraphics[width=0.49\textwidth]{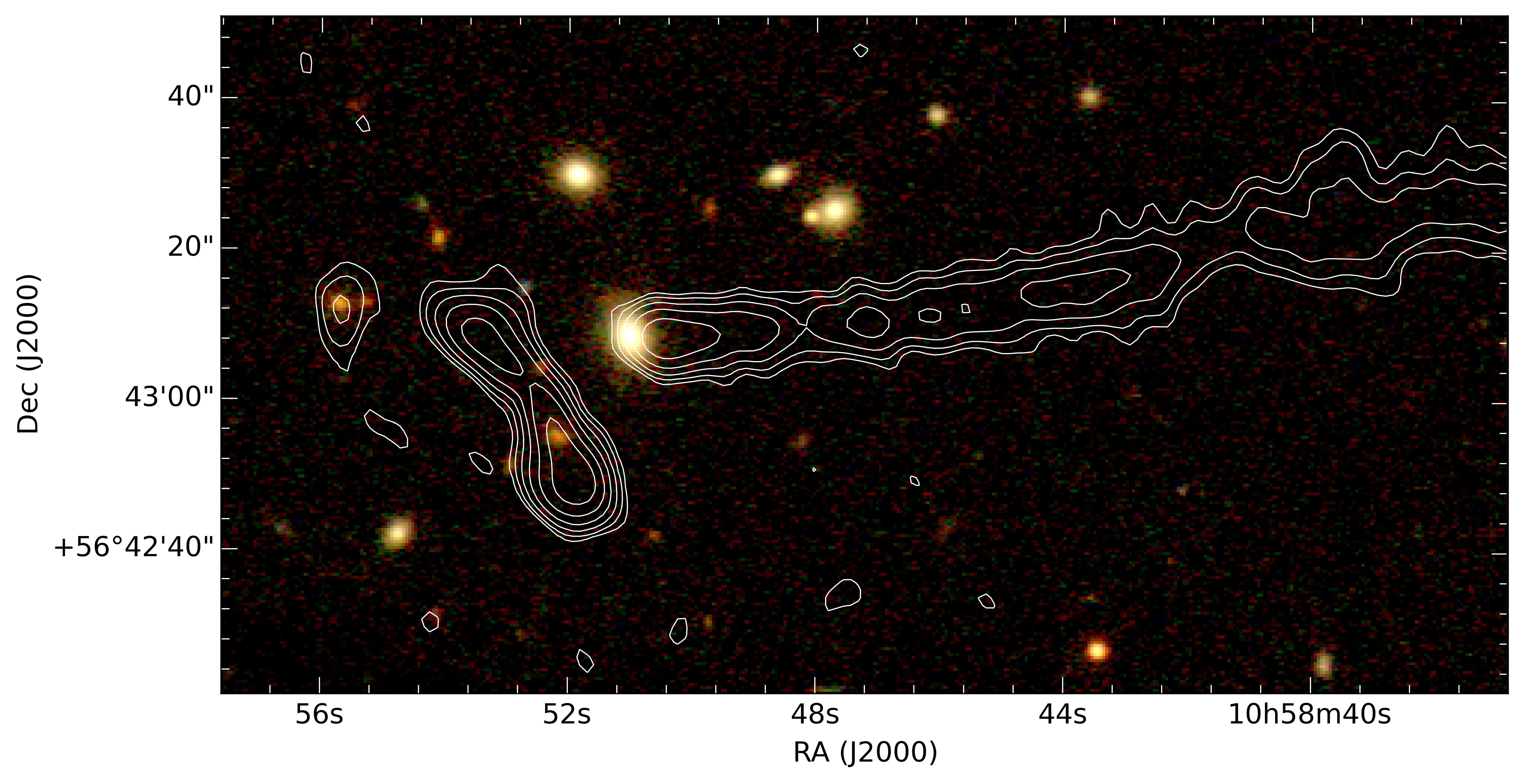}
  \includegraphics[width=0.49\textwidth]{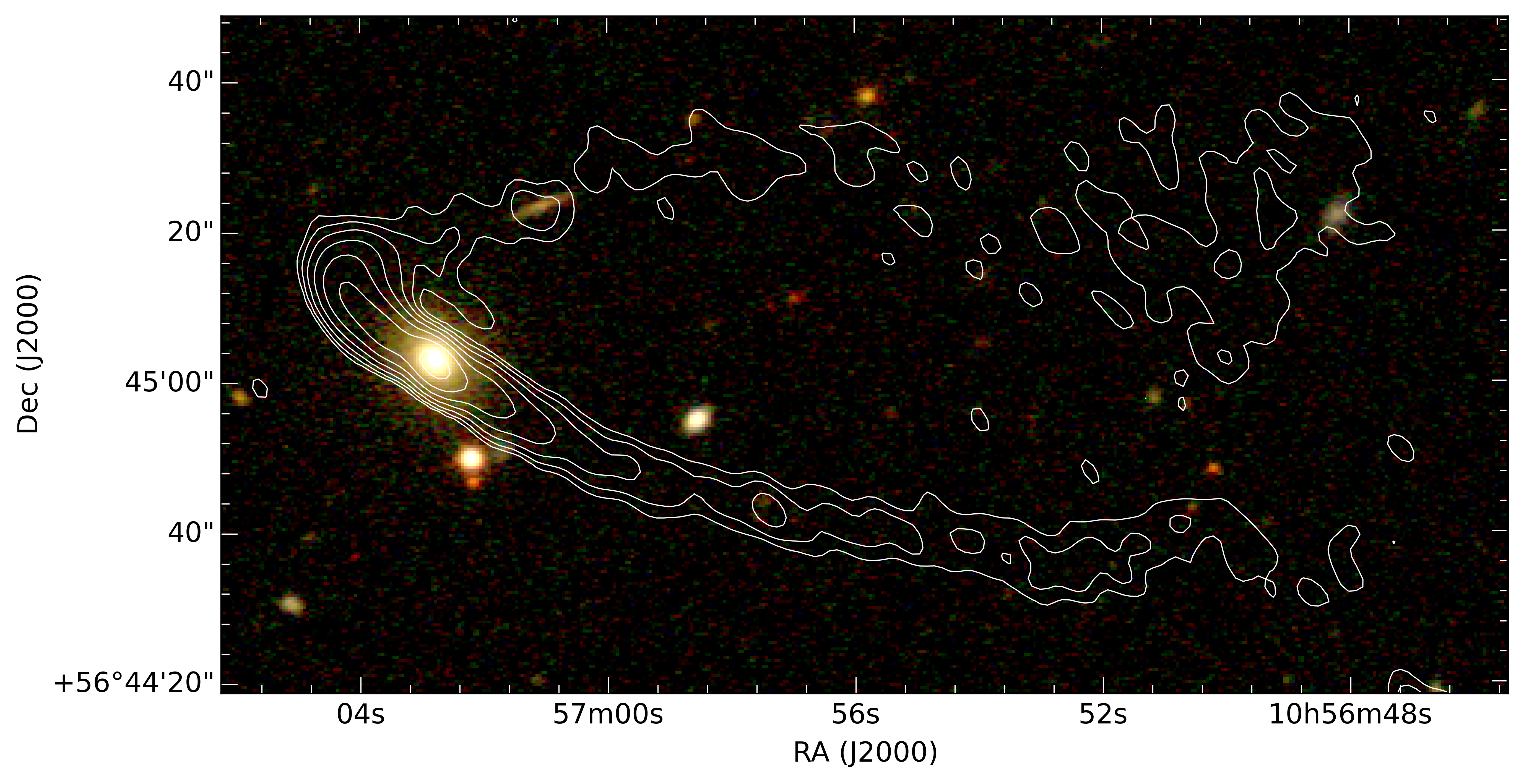}
     \caption{SDDS \textit{g, r, i} optical images overlaid with GMRT contours at 610 MHz. The levels are $[3,   6,  12,  24,  48,  96]\times\sigma$ where $\sigma = 20~\mu$Jy beam$^{-1}$. Left: the optical source associated with the HT GRG is the elliptical galaxy SDSS J105851.01+564308.5 at a redshift of $z \approx 0.139$, within Abell 1132. The double source to the east of the HT is likely associated with the background galaxy SDSS J105852.18+564255.3 at a redshift of $z \approx 0.496$. Right: the optical source associated with the WAT is the elliptical galaxy SDSS J105702.79+564503.1 at a redshift of $z\approx0.136$, within Abell 1132.  \label{HT}}
\end{figure*}

Observations of Abell 1132, as part of LoTSS Tier-1 \citep{2017A&A...598A.104S}, reveal several previously unknown regions of radio emission associated with the cluster. In Fig.~\ref{Abell1132-pr} we present our overview image of Abell 1132 where LOFAR radio emission in high- and low-resolution and Chandra X-ray emission are overlaid on an optical image from the Sloan Digital Sky Survey (SDSS). The cluster hosts diffuse radio emission near the cluster center, slightly offset from the X-ray emission, and a giant HT/NAT radio galaxy south of the cluster center with collimated emission to the east and diffuse emission to the west. A WAT radio galaxy, lying $\sim 0.2^{\circ}$ (1.8 Mpc) west of the cluster center, is also visible in our LoTSS observation (as shown in Fig. \ref{lo-contours}). Follow-up GMRT observations also show a significant portion of the HT radio galaxy and the full WAT radio galaxy at 325 and 610 MHz (see Fig.~\ref{325}). The heads of the two tailed radio galaxies coincide with elliptical galaxies belonging to the cluster: SDSS J105851.01$+$564308.5 at $z \approx 0.139$ (HT) and SDSS J105702.79$+$564503.1 at $z \approx 0.136$ (WAT) (see Fig.~\ref{HT}; redshifts taken from the Sloan Digital Sky Survey Data Release 2 from \citealp{2004AJ....128..502A}).

\subsection{Radio Halo Emission in Abell 1132}\label{3.1}

\begin{figure*}
\centering
\includegraphics[width=\textwidth]{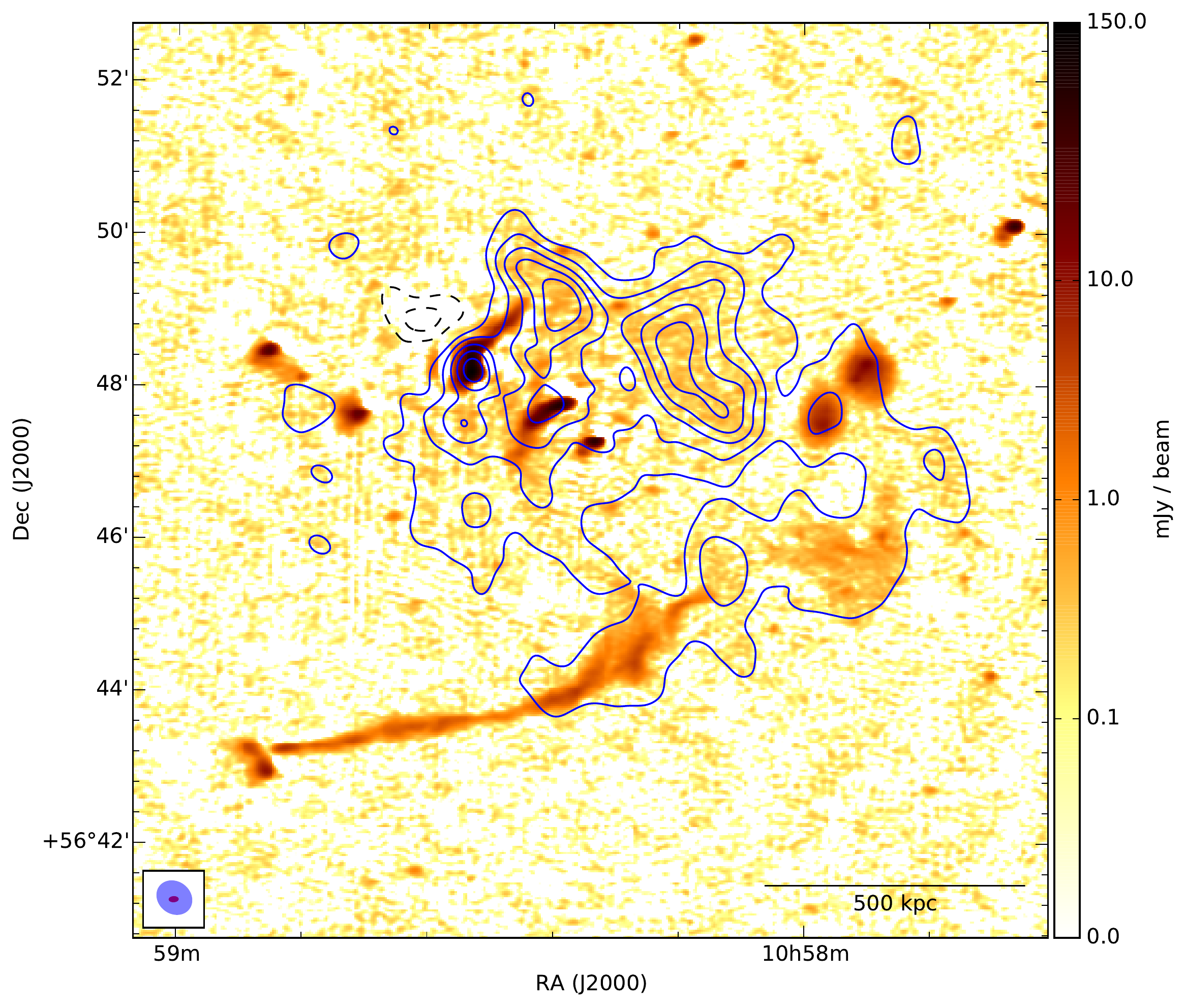}
\caption{Our FACTOR image: LOFAR high-resolution emission is shown in color (on a logarithmic scale), imaged with a beam size of $\sim8\arcsec\times5\arcsec$ and RMS noise of $120~\mu$Jy beam$^{-1}$, with LOFAR low-resolution diffuse emission contours overlaid in blue. LOFAR diffuse emission is imaged in CASA {\tt CLEAN} with an outer $uv$-taper of $20\arcsec$ and Briggs' robust 0 after subtracting compact sources imaged above a $uv$-range of 1000$\lambda$ (as explained in Sec.~\ref{3.1}). RMS noise of LOFAR low-resolution diffuse emission is $\sigma = 350\mu$Jy beam$^{-1}$ and the contour levels are $[-6, -3, 3, 6, 9, 12, 15]\times\sigma$. Beam size is designated by the red (high-resolution $\sim8\arcsec\times5\arcsec$) and blue (low-resolution $\sim30\arcsec\times26\arcsec$) ellipses.  \label{factor-subtracted}}
\end{figure*}

Cluster-scale diffuse emission, characteristic of a radio halo, is visible in both the LOFAR image at 144 MHz and the GMRT image at 325 MHz of Abell 1132. Fig.~\ref{lo-contours} shows our low-resolution LOFAR image with our low-resolution GMRT 325 MHz image contours overlaid. In Fig.~\ref{lo-contours} the halo appears more extensive in the LOFAR image, extending toward and possibly connecting to the diffuse emission of the giant southern HT galaxy. \\

\begin{figure*}
\centering
\includegraphics[width=0.49\textwidth]{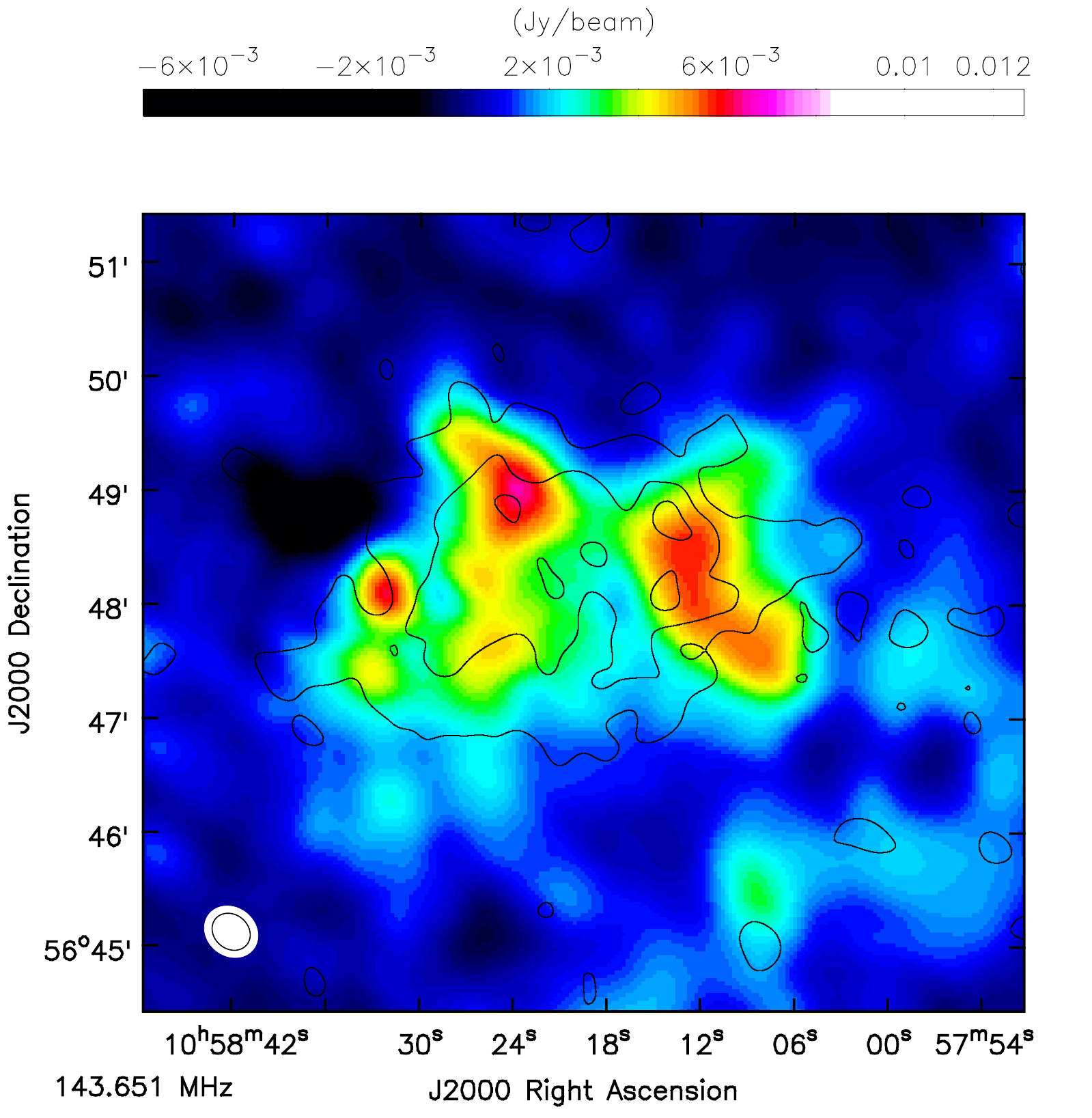}
\includegraphics[width=0.49\textwidth]{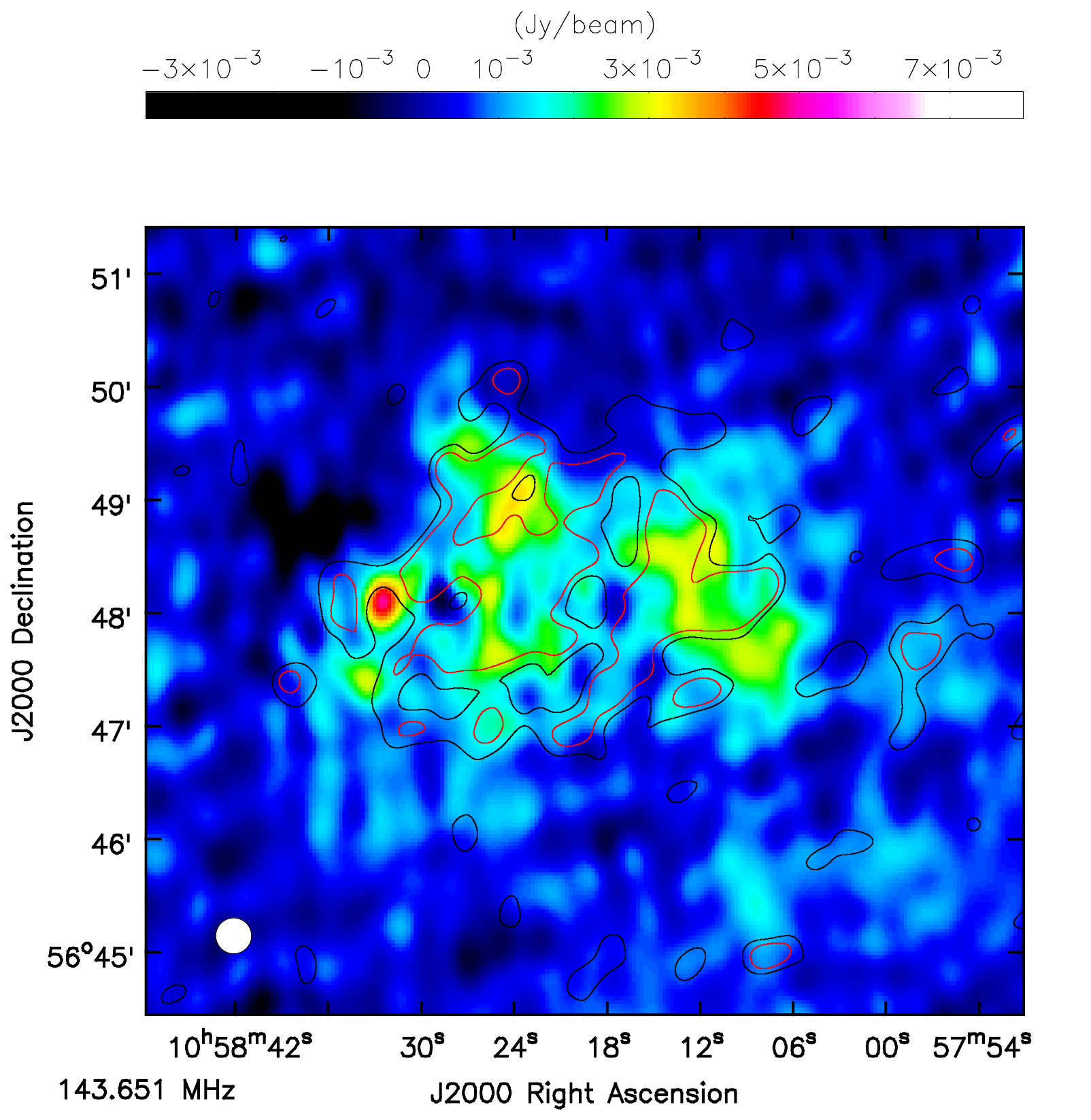}
\caption{LOFAR low-resolution diffuse emission after subtraction of compact sources (as explained in Sec.~\ref{index}) is shown in color and GMRT 325 MHz low-resolution diffuse emission after subtraction of compact sources is show as contours overlaid. Left: Our optimal image of the diffuse component in Abell 1132. Both LOFAR and GMRT compact-source-subtracted datasets were imaged in CASA {\tt CLEAN} with an outer $uv$-taper of 20\arcsec and Briggs' robust 0. Contours represent $[3, 6, 9]\times\sigma$ at 325 MHz where $\sigma = 75 ~\mu$Jy beam$^{-1}$. Right: Our uniform-weighted image of the diffuse emission in Abell 1132. Both LOFAR and GMRT compact-source-subtracted datasets were imaged in CASA {\tt CLEAN} with the same minimum $uv$-range ($100\lambda$), same outer $uv$-taper (20\arcsec), and uniform weighting, and were re-gridded and smoothed to the same beam (19\arcsec). Black contours represent 2$\sigma$ and red contours represent 3$\sigma$ at 325 MHz where $\sigma = 130~\mu$Jy beam$^{-1}$. The spectral index estimates stated in Sec.~\ref{index} were calculated by comparing the measured flux density within the 2$\sigma$ and 3$\sigma$ regions at 325 MHz. The red 3$\sigma$ contours at 325 MHz define an east and west region of the halo, where separate measurements were taken. The final estimate for the spectral index is an average of the values calculated within the 2$\sigma$ region and the two 3$\sigma$ regions. The residual flux of the brightest cluster galaxy was not included as measured flux in these regions.\label{halo-best}}
\end{figure*}

As seen in Fig.~\ref{lo-contours}, there are several bright and extended FRI galaxies near the cluster center. We performed a compact-source-subtraction on our 144 and 325 MHz data to better image the diffuse emission and eliminate contamination from the central radio galaxies. Since the halo is detected as diffuse emission on the scale of $\sim500 - 700$ kpc, we subtracted compact sources corresponding to emission spanning less than 500 kpc. At Abell 1132's redshift, this corresponds to visibility data greater than 1000$\lambda$ in the $uv$ plane. We made an image in CASA {\tt CLEAN} with a $uv$-cut below 1000$\lambda$ and an outer $uv$-taper of $10\arcsec$, and subtracted the model component from the $uv$ data using CASA tools {\tt FT}\footnote{We used our own CASA task called {\tt FTW} which includes the widefield w-projection parameter.} and {\tt UVSUB}. We then re-imaged the source-subtracted datasets with their full $uv$-range\footnote{\textgreater~80$\lambda$ for 144 MHz and \textgreater~100$\lambda$ for 325 MHz} to bring out extended emission (see Fig.~\ref{halo-best}: Left). The residual emission from the central galaxies is < 1\% of their flux densities. \\

\begin{figure}
\centering
\includegraphics[width=0.49\textwidth]{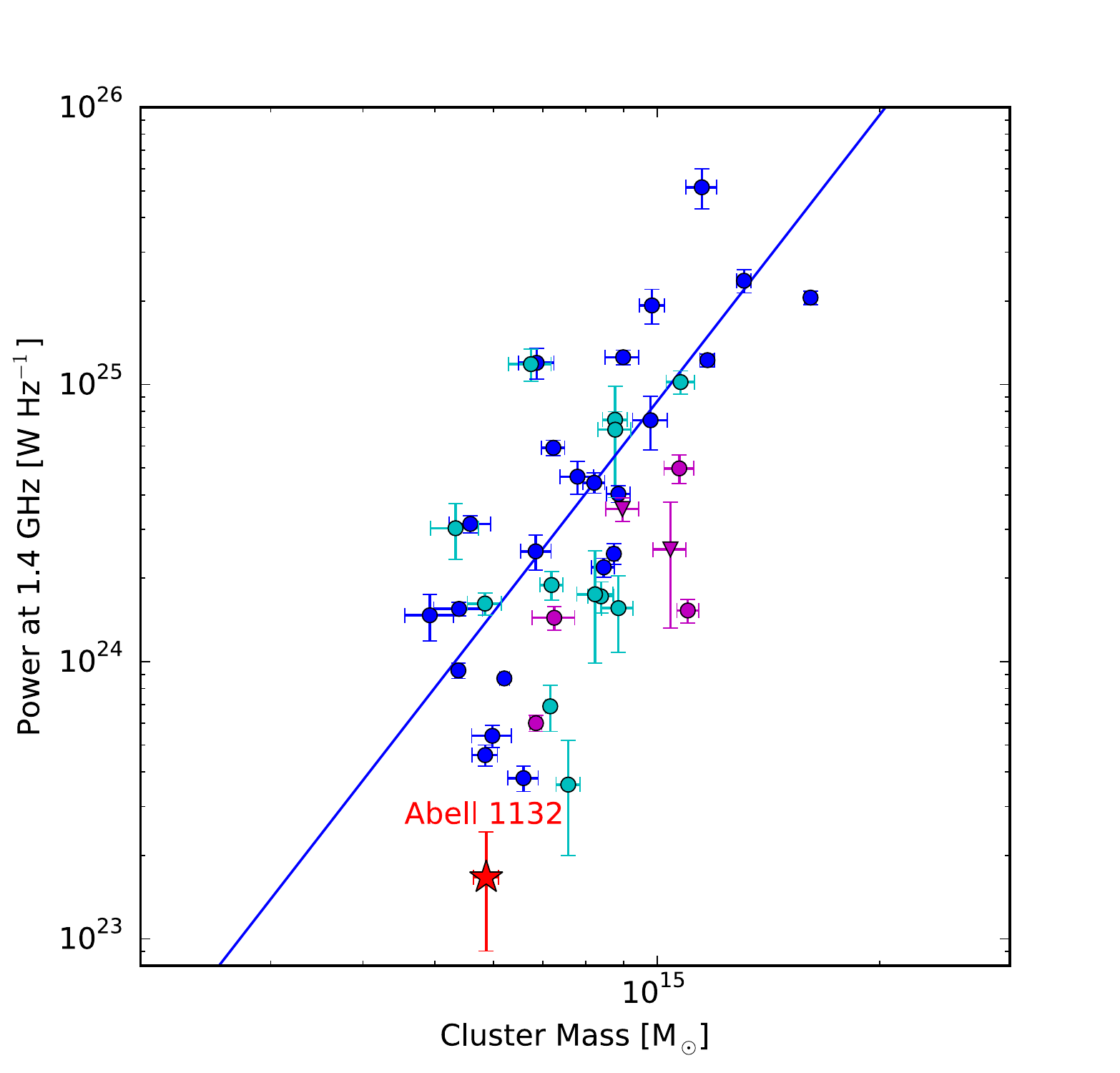}
\caption{A sample of radio halos plotted by their radio power at 1.4 GHz versus their cluster mass (M$_{500}$ -- as determined from Planck observations). The sample of halos and their correlation is reproduced from \citet{2016A&A...595A.116M}. Halos with flux measured at 1.4 GHz are marked by blue circles and their derived fit is shown as a blue line. Cyan circles represent halos with flux measured at frequencies other than 1.4 GHz. Magenta circles represent ultra-steep halos, and magenta triangles represent ultra-steep halos with flux measured at frequencies other than 1.4 GHz. Abell 1132 is marked by the red star, and falls well below the correlation line as well as below all the halos in this sample. \label{corr}}
\end{figure} 

LOFAR diffuse emission, after compact-source-subtraction, is shown in red in Fig.~\ref{Abell1132-pr} and as contours overlaid on our FACTOR image in Fig.~\ref{factor-subtracted}. It is apparent in Fig.~\ref{factor-subtracted} that there is a hole (negative artifact) to the north-east of the brightest cluster galaxy. The negative artifact likely occurred because there was imperfect calibration, modelling, and subtraction of the brightest cluster galaxy, as it has no prior model and it is embedded in diffuse halo emission. In LOFAR facet-calibrated images, negative artifacts, or negative bowls, often occur near bright sources that have not been previously modeled. This hole appears to be the only one in the cluster field and is on a relatively small scale. The negative artifact remains after the compact-source-subtraction, and is indicated by negative contours (dashed) in Fig.~\ref{factor-subtracted}. It is possible that the halo emission extends within this region, but future low-frequency observations would be needed for confirmation.  \\

Diffuse emission, after compact-source-subtraction, at 144 and 325 MHz are shown together in Fig.~\ref{halo-best}. The size and morphology of the halo is comparable in both the LOFAR and GMRT images, however, the LOFAR image of the halo shows some additional, weaker emission to the south, possibly connecting to the diffuse emission in the giant HT. The bulk of the detected diffuse emission lies slightly west of the cluster center, exhibiting a subtle offset ($\sim 200$ kpc) from the center of the Chandra X-ray emission. The halo takes on a roughly elliptical shape, shorter in the north-south direction and elongated from south-east to north-west, with a major axis of $\sim$ 750 kpc and minor axis of $\sim$ 570 kpc, as seen by LOFAR. The surface brightness of the halo (after subtraction of central galaxies) within $3\sigma$ contours is $\sim 0.8~\mu$Jy arcsec$^{-2}$ at 144 MHz with $\sigma = 350~\mu$Jy beam$^{-1}$ and a beam size of $30 \arcsec\times26\arcsec$ (see Fig.~\ref{factor-subtracted}). \\

We produced GMRT images at 610 MHz tapered to $30\arcsec$ resolution to enhance diffuse emission, but the radio halo is not detected above 2$\sigma$ where $\sigma \approx 100~\mu$Jy beam$^{-1}$. Archival VLA observations performed in D array were retrieved, reduced and re-imaged, but diffuse emission near the cluster center is not detected above 2$\sigma$ where $\sigma \approx 250~\mu$Jy beam$^{-1}$. \\

\subsubsection{Halo spectral index estimate}\label{index}
We estimate the spectral index of the halo by imaging the compact-source-subtracted datasets at 144 and 325 MHz  in CASA {\tt CLEAN} with the same minimum $uv$-range ($100\lambda$), same outer $uv$-taper (20\arcsec), and uniform weighting, and compare the flux densities within the same region after re-gridding\footnote{\label{offset}We shifted the LOFAR map by -2 pixels in X-direction and -3 pixels in Y-direction to correct for an astrometric offset. This pixel shift was determined by comparing the high-resolution LOFAR and GMRT maps imaged with the same settings and convolved to the same beam (11\arcsec) and calculating the offset of the maximum pixel of several point sources near the cluster center.} and smoothing to the same beam size ($19\arcsec\times19\arcsec$). In a region indicated by $2\sigma$ contours at 325 MHz where $\sigma = 130~\mu$Jy beam$^{-1}$, the spectral index is $\alpha=-1.80\pm0.18$. In a region indicated by $3\sigma$ contours at 325 MHz (east portion of halo), the spectral index is $\alpha=-1.71\pm0.19$, and in another region indicated by $3\sigma$ contours at 325 MHz (west portion of halo) the spectral index is $\alpha=-1.74\pm0.20$. (See the regions as contours in Fig.~\ref{halo-best}: Right.) Therefore, we give an average spectral index estimate of $\alpha=-1.75\pm0.19$ and classify this radio halo as ultra-steep. With a spectral index of $\alpha=-1.75$, the surface brightness of the halo emission would be $\sim1.3~\mu$Jy arcsec$^{-2}$ at 1.4 GHz, and considering a 15\% error\footnote{We approximate the error in the halo's flux by assuming a 10\% error from FACTOR calibration, modeling, and imaging (based on experience) and introducing a 5\% error from the contamination of the residual emission of subtracted central galaxies.} in our total measured flux at 144 MHz and an error of $\pm0.19$ in the spectral index, the radio power at 1.4 GHz is determined to be P$_{1.4}=(1.66\pm0.76)\times10^{23}$ W Hz$^{-1}$. \\

It has been found that the radio power of halos correlate with the X-ray luminosity of the host cluster \citep{2007ApJ...670L...5B, 2009A&A...507..661B, 2013ApJ...777..141C, 2015ApJ...813...77Y}. In Fig.~\ref{corr}, we plot the radio power at 1.4 GHz versus the Planck cluster mass $M_{500}$ for a sample of radio halos and include Abell 1132's halo, indicated by the red star. The halo is not only ultra-steep but also extremely weak: the plot in Fig.~\ref{corr} shows Abell 1132's halo lying well below the correlation line. It is possible that this ultra-steep halo sets an unprecedented record for the weakest halo discovered so far. The fact that it is so steep and weak, as seen at low-frequency, is consistent with the non-detection of diffuse emission at 1.4 GHz.

\subsection{Giant Radio Galaxy: Head-Tail }
The GRG is a prominent feature of the radio emission from the cluster, exhibiting long and narrow emission with a projected linear size of 1.3 Mpc. The head of this emission coincides with an elliptical galaxy$^{\ref{opticalgalaxyHT}}$ near the same redshift as Abell 1132 (see Fig.~\ref{HT}). The giant tail, as seen by LOFAR, is the same head-tail source as seen in NVSS and in reprocessed WENSS data, but there the tail is only detected to be 370 kpc long \citep{Rudnick2009}. The GRG appears to be one-sided, since only one jet is visible. It is likely that the jets have joined into a single tail aligned behind the trajectory of the galaxy as it has moved west-to-east. The giant tail in Abell 1132 has similar physical characteristics to the tail in Abell 2256 \citep{owen2014}, however, it is more than twice as long as the tail in Abell 2256. If the host galaxy was moving at the sound speed of the cluster, $\sim 1000$ km s$^{-1}$, then it is possible that the AGN has been active for $\sim1$ Gyr, which is much greater than typical AGN life cycles (on the order of a few Myr).\\

The GRG has traveled from west-to-east through the cluster outskirts, leaving the observed tail as a trail of AGN emission, but its vast extent challenges the fact that the tail electrons 1 Mpc and further from the AGN head should no longer be emitting. The separation in the tail at 1 Mpc, where the emission becomes more diffuse (see Fig.~\ref{Abell1132-pr} \& \ref{factor-subtracted}), is interesting: here, dormant tail electrons may have been disrupted and re-accelerated, leading to a re-brightening that gives rise to the additional 300 kpc of diffuse emission (\citealp{2017degasperin} reports a similar re-brightening of dormant tail electrons from a WAT radio galaxy within the massive merging cluster Abell 1033 and attributes it to ``gentle re-energisation''). It also appears that the radio halo connects to the diffuse portion of the tail (see Fig.~\ref{factor-subtracted}). A connection between the halo and tail may give an indication of where the seed particles needed for turbulent re-acceleration come from. However, a question remains in how the length of the tail up to 1 Mpc could retain its collimated form if it has been affected by merger turbulence. Since the diffuse portion is not visible in the GMRT images at 325 or 610 MHz (Fig.~\ref{325}), it is likely to be very steep and very weak emission. If the diffuse portion has been re-accelerated, its emission should have a slightly flatter spectral index than the steepest part of the collimated portion. \\

\begin{figure}
\centering
\includegraphics[width=0.6\textwidth]{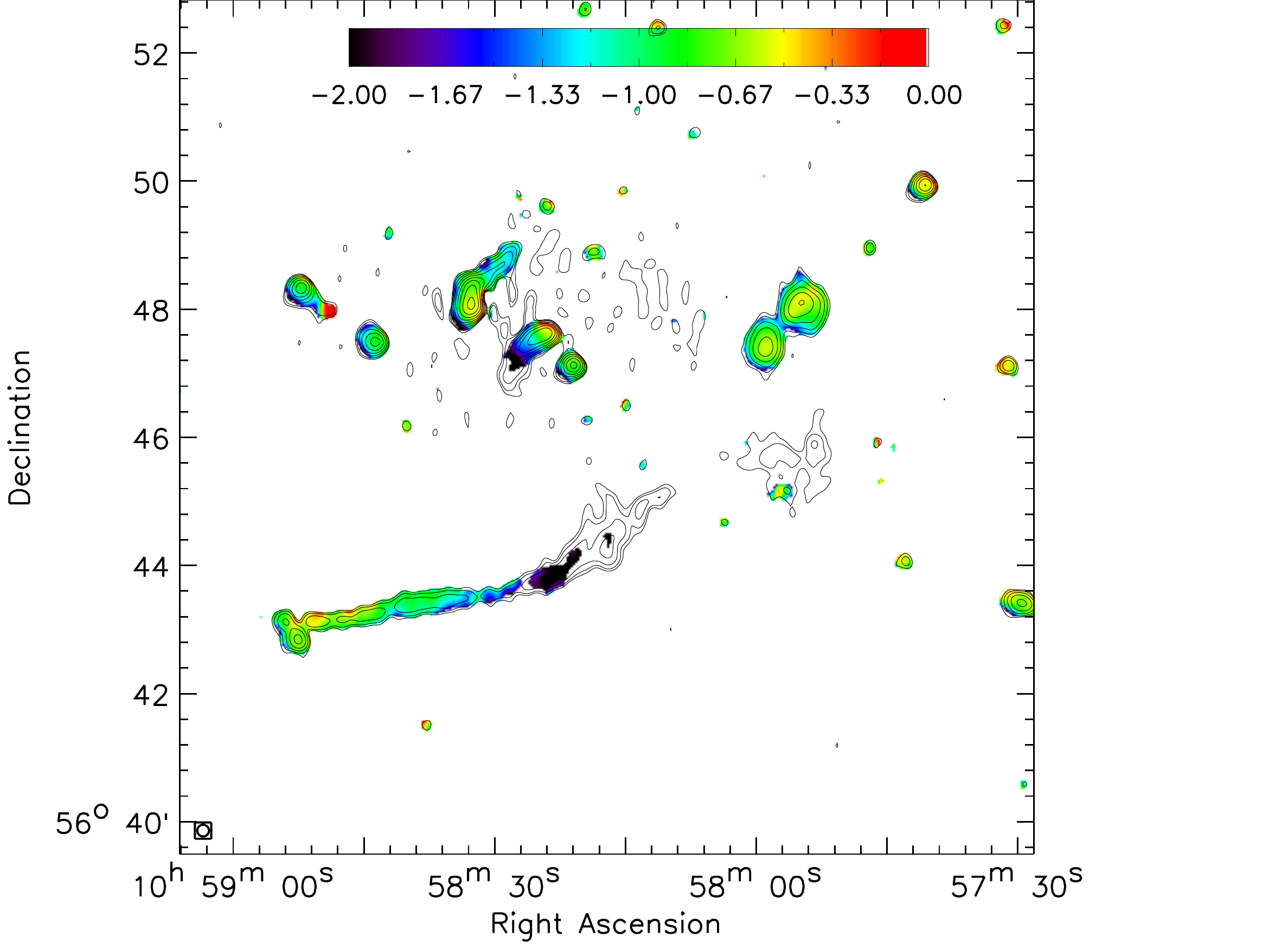}
\caption{Spectral index map of the giant HT galaxy using emission cut above 3$\sigma$ from LOFAR at 144 MHz and GMRT at 325 and 610 MHz. LOFAR and GMRT emission are imaged in CASA {\tt CLEAN} with an outer $uv$-taper of 10$\arcsec$, uniform weighting, and a minimum $uv$-range of $200 \lambda$ with RMS noise of $\sigma_{140} = 300~\mu$Jy beam$^{-1}$, $\sigma_{325} = 100~\mu$Jy beam$^{-1}$, and $\sigma_{610} = 60~\mu$Jy beam$^{-1}$. All images were smoothed to the same beam size ($11\arcsec\times11\arcsec$). Maps at 144 MHz and 325 MHz were re-gridded to the map at 610 MHz$^{\ref{offset}}$. LOFAR emission is also shown as black contours with levels $\sigma_{140}\times[ 3.45, 6, 12, 24, 48, 96, 192, 384, 768]$. \label{tail-spectral}}
\end{figure}

In Fig.~\ref{tail-spectral} a spectral index map of LOFAR and GMRT images aids in classifying the HT galaxy. The spectrum steepens along the length of the tail, with a spectral index of $\alpha\approx -2$ at a distance of $\sim800$ kpc from the head. At further distances only LOFAR detects emission, hence a spectral index could not be determined for the detached portion of the tail beyond 1 Mpc. An upper limit can be placed for the detached, more diffuse portion by comparing the mean flux at 144 MHz to $2\sigma$ at 325 and 610 MHz (see caption of Fig.~\ref{tail-spectral} for values of $\sigma$). The upper limit of the spectral index is $\alpha\textless-2.3$ for 325 MHz and $\alpha\textless-1.7$ for 610 MHz.  \\

The WAT to the west of the cluster center (see Figs.~\ref{lo-contours} \& \ref{325}) follows a similar trajectory to the HT. It appears to be moving from west-to-east on the outskirts of the cluster, leaving two tails of AGN emission. The longer tail has a projected size of $\sim650$~kpc and the shorter tail has a projected size of $\sim450$~kpc.

\section{Discussion and Conclusions} \label{discuss}

The discovery of this radio halo in Abell 1132 demonstrates LOFAR's potential to detect weak, steep-spectrum emission on large scales. The steep-spectrum halo in Abell 1132 is noteworthy for three reasons: \\
\begin{enumerate}[(i)]
\item with a spectral index of $\alpha=-1.75\pm0.19$ it is one of the steepest halos detected to date\\
\item with a size of $\sim$~650 kpc and radio power of P$_{1.4}=(1.66\pm0.76)\times10^{23}$ W Hz$^{-1}$ the halo is smaller than usual and it is remarkably faint\\
\item the halo is $\sim$~200 kpc offset from the X-ray emission of the cluster.\\
\end{enumerate}

A few other ultra-steep halos include $\alpha \approx -2.1$ in A521 \citep{2008Natur.455..944B}, $\alpha \approx -1.7 to -1.8$ in A697 \citep{2010A&A...517A..43M}, and A1682 \citep{2013A&A...551A.141M}. Steep-spectrum radio halos challenge hadronic models because energy arguments rule out the possibility that very steep halos with $\alpha \sim -1.5 to -2$ are produced by cosmic-ray electrons that follow power-laws in momentum \citep{2004A&A...413...17P, 2008Natur.455..944B}. However, turbulent re-acceleration models do predict a large population of steep-spectrum halos, which exhibit a break in their electron spectra near energies of a few GeV \citep{2004MNRAS.350.1174B, 2008Natur.455..944B}. Most radio halos known today, discovered at higher radio frequencies (1.4 GHz), are produced by mergers between the most massive galaxy clusters. According to the turbulent re-acceleration model, radio halos with much steeper spectra should be produced by less energetic, more frequent mergers. Hence, the bulk of radio halos may have yet to be discovered because they are only visible at low frequencies. Given its sensitivity to diffuse emission with low surface brightness, LOFAR will be a valuable tool to reveal this population.  \\

In hadronic models the radio emission of the halo should also roughly follow the X-ray surface brightness, as the X-ray emission traces the thermal ICM that provides the targets for the hadronic collisions. Clearly, this is not the case in the halo in Abell 1132. The fact that the halo is smaller and fainter also supports a turbulent re-acceleration origin of halos, as the steeper-spectrum halos are produced by older populations of relativistic electrons that can no longer sustain a luminous halo. Hence, we may be witnessing a radio halo that is transitioning into an ``off'' state. It would be interesting to determine the dynamical status of the cluster merger using optical spectroscopy, in order to relate the radio properties of the halo to the phase and energetics of the cluster merger. With much better spectral capabilities than the current X-ray telescopes, the future X-ray observatory ATHENA may be able to probe turbulence within Abell 1132's intracluster medium.\\

The discovery of the radio halo and GRG found together in one cluster has raised the question about a possible connection. The presence of a radio halo suggests that the cluster has recently undergone a merger, but it is unclear whether the merger has affected the emission of the GRG. Near the GRG-head the emission appears to be mostly undisturbed, very narrow, and collimated, but the furthest emission of the tail is more diffuse. It may be possible that the 1.3 Mpc tail is visible because it has been re-accelerated by turbulent merger activity. A spectral index measurement of this disturbed portion as compared to the collimated portion would reveal the age of this emission and clarify whether it has been re-accelerated. This cluster is a prime target for studying merger mechanisms and the re-acceleration of dormant electrons. \\

\section*{Acknowledgements}

This work was supported by the Deutsche Forschungsgemeinschaft (DFG) through the Collaborative Research Centre SFB 676 ``Particles, Strings and the Early Universe", project C2. LOFAR, the Low Frequency Array designed and constructed by ASTRON, has facilities in several countries, that are owned by various parties (each with their own funding sources), and that are collectively operated by the International LOFAR Telescope (ILT) foundation under a joint scientific policy. The LOFAR software and dedicated reduction packages on \url{https://github.com/apmechev/GRID_LRT} were deployed on the e-infrastructure by the LOFAR e-infragroup, consisting of J. B. R. Oonk (ASTRON \& Leiden Observatory), A. P. Mechev (Leiden Observatory) and T. Shimwell (Leiden Observatory) with support from N. Danezi (SURFsara) and C. Schrijvers (SURFsara). This work has made use of the Dutch national e-infrastructure with the support of SURF Cooperative through grant e-infra160022. We thank the staff of the GMRT that made these observations possible. GMRT is run by the National Centre for Radio Astrophysics of the Tata Institute of Fundamental Research. This research has made use of the NASA/IPAC Extragalactic Data Base (NED) which is operated by the JPL, California institute of technology under contract with the National Aeronautics and  Space administration. TS acknowledges support from the ERC Advanced Investigator programme NewClusters 321271. F.A-S. acknowledges support from {\em Chandra} grant GO3-14131X. EKM acknowledges support from the Australian Research Council Centre of Excellence for All-sky Astrophysics (CAASTRO), through project number CE110001020. RM gratefully acknowledges support from the European Research Council under the European Union's Seventh Framework Programme (FP/2007-2013) ERC Advanced Grant RADIOLIFE-320745. AOC gratefully acknowledges support from the European Research Council under grant ERC-2012-StG-307215 LODESTONE. We also thank A. Botteon (Università di Bologna) and G. Kokotanekov (University of Amsterdam) for their helpful comments.




\bibliographystyle{mnras}
\bibliography{Wilber-et-alA1132} 








\bsp	
\label{lastpage}
\end{document}